\documentclass[12pt]{article}
\usepackage{fullpage}

\def\be{\begin{equation}}
\def\ee{\end{equation}}

\begin{document}

\begin{flushright}
TIFR/TH/08-18
\end{flushright}
\bigskip

\begin{center}
\Large{\bf Bohm's realist interpretation of Quantum mechanics} \\
\bigskip\bigskip
\large{Virendra Singh} \\
\bigskip
INSA C.V. Raman Research Professor \\
Tata Institute of Fundamental Research \\
1, Homi Bhabha Road, Mumbai 400 005, India 
\end{center}
\bigskip

\begin{center}
\underbar{Abstract}
\end{center}
\bigskip\bigskip

A brief account of the world view of classical physics is given
first.  We then recapitulate as to why the Copenhagen interpretation
of the quantum mechanics had to renounce most of the attractive
features of the clasical world view such as a causal description,
locality, scientific realism and introduce a fundamental distinction
between system and apparatus.  The crucial role is played in this by
the Bohr's insistence on the wavefunction providing the most complete
description possible for an even individual system.  The alternative of
introducing extra dynamical variables, called hidden variables, in
addition to the wavefunction of the system so as to be able to retain
at least some of the desirable features of classical physics, is then
explored.  The first such successful attempt was that of Bohm in 1952
who showed that a realistic interpretation of the quantum mechanics
can be given which maintains a causal description as well as does not
treat systems and measuring appeartus differently.  We begin with the
construction of the Bohm's theory.  He introduces particle positions
as the hidden variables.  The particle positions play a special role
in Bohm theory.  The particle trajectories are guided by the
wavefunction.  The Bohm theory is deterministic.  The probability
enters through a special assumption, ``quantum equilibrium''
hypothesis, for the initial conditions on the ensemble of particle
trajectories.  The ``wave or particle'' dilemma is resolved by a
``wave and particle'' resolution.  The measurements 
in Bohm theory can be described without mysticism.  Bohm's theory is
however nonlocal.  It is however without nonlocal signalling.  After Bell's
work, and the experimental work on testing Bell's inequalities, it has 
however, become clear that quantum mechanics is basically
nonlocal.  We also describe briefly the ``Bohmian mechanics''
reformulation of the Bohm theory.  In the end we discuss some
discontents with the Bohm's theory as well as it's future
prospects. The writeup is supplemented with mathematical and
bibiliographical notes.

\newpage

\noindent {\large\bf 1. The World view of Classical Physics:}
\bigskip

The description of the physical universe, as given in classical
physics, was in many ways a very attractive one.  It firmly subscribed
to scientific realism.  It aimed for internal logical consistency and
completeness of description.  As it described the world as it is, it
was very satisfying.
\bigskip

The basic ontological entities were point-particles, fields and
space-time.  They obeyed the well defined causal dynamical laws having
the form of differential equations and were deterministic.  Thus in
Newtonian mechanics if the positions and momentums of all the particle
were specified at any one time, their motion could be determined for
all times.  Similarly Maxwell's equations for electromagnetic field
and Einstein's equations for gravitational (-metric) field were
determined if they were specified for any one time together with
appropriate boundary conditions.  There was no fundamental role for
randomness in it's physical description.  The role of the probability
consideration was thus only present when either we were not interested
in full details of the situation and we wanted to have a simplified description
of a complex situation using only a few variables.  Such examples are
provided by classical statistical mechanics and the theory of Brownian
motion.  With the discovery of chaotic nonlinear systems in classical
physics one has now to make a distinction between determinism and
predictability.  For these systems there is an extreme sensitive
dependance of the dynamical motion on initial conditions.
\bigskip

In it's mature form classical physics also shuns ``action at a
distance'' theories of influence.  This is achieved through the
modalities of fields.  Particles generate fields which then act on
other particles elsewhere.  As Faraday said ``matter can not act where
it is not''.  The physical effects and signals can not propagate
instantaneously.  They can do so at most with the speed of light.
\bigskip

Another aspect of classical physics, which we have come to admire more
in the post classical physics days is it's unitary nature.  Both the
physical systems and the measuring apparatus used to study them obey
laws of classical physics.  The measurement does not constitute an
epistomological problem.  Of course every measuring apparatus used to
probe a system will disturb it somewhat and we would be learning about
the disturbed system.  But in classical physics the disturbance can be
reduced to be as small as we like by using gentler probes.
\bigskip\bigskip

\noindent {\large\bf 2. Coming of the Quantum:}
\bigskip

As is well known this beautiful edifice of classical physics, after
successfully serving for the description of macroscopic physical world
since Newton till Einstein, ie from the seventeenth century to the
nineteenth century, was found empirically inadequate in microscopic
world of atoms and radiation.  The first quarter of the twentieth
century was the period of struggle for the new quantum ideas.  The
final mathematical formulation quantum mechanics, needed to describe
new phenomenon in the microscopic domain, was finally achieved around
1925. 
\bigskip

Soon thereafter the ``Copengahen interpretation'' of what the new
mathematical quantum formalism means emerged.  It became the ruling
orthodoxy for a long time so that any other interpretation of the
formalism was not encouraged to get a foothold.  In the Copenhagen
interpretation the scientific realism, the bedrock of the classical
physics, was given up along with determinism, unitaryness and many
other appealing features of classical physical description.  That a
realist interpretation of quantum mechanics could be given by realised
by David Bohm and published in 1952. A precursor was de Broglie's  
attempt called ``pilot wave'' interpretation given in 1927.  In view
of the dominate of the 
Copenhagen interpretation, it was however not taken seriously until
the important work of John Bell on foundations of quantum mechanics in
late nineteen sixtees. 
\bigskip

In order to put things in perspective and bring out the magnitude of
Bohm's achievement in proposing his realist interpretation, we will
first understand as to why Copenhagen interpretation was forced to
renounce so many of appealing aspects of classical physics.  As a
further preliminary we now present a brief account of the ``rules of
game'' on which all interpretations of quantum mechanics agree in
order to make use of the mathematical formalism.
\bigskip\bigskip

\noindent {\large\bf 3. Rules of the Game:}
\bigskip

\begin{enumerate}
\item[{(i)}] \underbar{The States}: The state of a system at any time
is described by a state vector.  A state vector, when multiplied by a
complex number, also describes the same state.  The coordinate
representative of the state vector, ie the wavefunction, is generally
a complex quantity.  If two different state vectors are appropriate to
describe the state of a system, then a linear combination of these two
also describes a possible state of the system.  This principle of
linear superpositions of state vectors is of the same nature as for
classical electromagnetic waves.  It is this feature of quantum
mechanics which helps explain the wave nature of quantum particles
like electrons in the electron interference experiments.  All state
vectors of a physical system belong to the Hilbert space of the
system. The state vectors will normally be assumed to be normalised to
unity. 
\item[{(ii)}] \underbar{Physical observables}: All physical
observables are represented by linear self-adjoint operators  
operating on the state vectors in the Hilbert space of the system.  All
eigenvalues of such operators are real numbers.  Any measurement
of an observable always results in getting one of it's eigenvalues.
The energy of the system is given by it's Hamiltonian operator.
\item[{(iii)}] \underbar{Dynamics}: The time evolution of a state
vector is described by a linear Schr\"odinger equation.  The
evolution is unitary with Hamiltonian acting as time-translation
operator.  If the system at any one time is prepared in any
particular state it fixes the state vector at any other later time
during it's free evolution ie before it is measured. \bigskip \\ All the
postulates so far are quite consistent with a determistic theory.  The
probability enters into the theory through the following rule.
\item[{(iv)}] \underbar{Statistical postulate}: As we noted earlier
the measurement of any observables results in only one of it's
eigenvalues being observed.  The probability of
any particular eigenvalue being observed is given by Born's rule ie it
is equal to absolute square of the component of the eigenvector
corresponding to the observed eigenvalue in the state-vector at the
time of measurement.
\end{enumerate}
\bigskip\bigskip

\noindent {\large\bf 4. Renunciations of Copenhagen interpretation:}
\bigskip

The Copenhagen interpretation was hammered out by Niels Bohr and
collaborators including Heisenberg, Pauli, Rosenfeld and others.
They had the difficult job of making some sense of puzzling quantum
phenomenon with which they had to struggle using the above rules of
the game.  There was also the problem of the nature of quantum entities.  The
light behaved as waves in some situation involving their interference
and diffraction, while it behaved as particles, called `photons', in
situations involving interaction of light with matter, such as
photoelectric effect and Compton effect.  Electrons were regarded as
charged point-particles when discovered by J.J. Thomson in 1897 but
electron beams were later seen to exhibit diffraction from crystals in
the experiments by Davisson and Germer in 1927 showing that they too
had a wave behavior.  Then there was the notorious problem, to which
we will come back, of how measurements of a physical observables of a
quantum systems produce \underbar{definite} answers.
\bigskip

Niels Bohr was quite ascetic in his attitude towards new concepts.  He
took it as bedrock the idea that the description of a quantum system
provided by the state vector was \underbar{complete} in itself for
even an individual system.  No more completer description was
possible.  It was not a statistical description of an ensemble of
similar systems as was advocated by Einstein.  All the renunciations
follow from this stand point.
\bigskip

From the quantum rules it follows that when we measure an observable for a
individual system which is in a super position of two eigenvalues of
this observable, we will obtain the result to be one of these two
eigenvalues.  We can not say which one it will be.  Born rule only
gives the probability for each of these two outcomes.  This clearly
leads to violation of causality and determinism if the wavefunction
provides a complete description of the system.
\bigskip

We next look at the double slit experiment with electrons.  Each
electron in the beam after it has passed through the slits is detected at
a single point on the screen in a detector.  It exhibits a
discreteness as expected from the point particles.  However the
vertical distributions of electron clicks, produced by the arrival of
electrons at the screen exhibits an interference pattern
indicating of the wavenature of the electron.  In Copenhagen view, as
quantum rules apply to each individual process, the interference
pattern is due to each electron interfering with itself.  It is not
due to an average produced by many electrons in an ensemble.
Classically electron must have gone either one or the other of the
slits.  It can then however not produce the interference pattern on
the screen. We could try to actually check as to through which slit a
particular electron went by putting an electron detector at each slit
and noting whether it clicks or not.  We then would know it's path, ie
the ``which way'' informations about it.  We then find that each
electron goes either one or the other slit only.  But now the
interference pattern is no longer seen.  What we see is a
superposition of two distributions, each one corresponding to
electrons coming from one of the slits.  It is just as we would get
for classical bullets.  So we get a point-particle like pattern when
we do know the ``which way'' information and a wave like pattern when
we do not know the ``which way'' information about the electrons. 
\bigskip

The moral Bohr would draw from the double slit experiment on electrons
that the phenomenon we observe are not produced by a physical system,
as it exists out there independently of us, but only through the
combined setup of physical system plus the appratus used for probing
it.  ``No phenomenon is a phenomenon until it is observed'' as Wheeler
puts it.  It thus forces us to renounce scientific realism for
quantum phenomenon.
\bigskip

For Bohr the measuring apparatus is described by classical physics
while the quantum system is to be described by the quantum rules.  For
him it is a logical necessity as the language of classical physics is
the only means to communicate the results of an experiments to each
other.  Thus in view of Bohr the description the quantum physics is
not unitary.  The system and the apparatus are not described using the same
framework as in classical physics.
\bigskip

John von-Neumann would rather have the measuring apparatus also
described using quantum dynamics.  He then has to introduce an
additional type of dynamics, not given by Schr\"odigner equation,
according to which every measurement, when completed, results in the
wavefunction of the system suddenly changing to the system being in
the eigenstate corresponding to the measured value of the observable.
This is the postulate of ``the collapse of the wavefunction'' at the
completion of the measurement.  The quantum rules of the game here are also
not enough since we have two kind of dynamics ie one applicable to
measurement interaction apart from the normal dynamics giving
Schr\"odinger equation.  The ``rules of game'' had included only the
Schr\"odinger dynamics in the dynamics.
\bigskip

Besides all the renunciations of scientific realism, causality,
determinism, unitary nature of system and apparatus description, it 
seemed that even the ``action at a distance'' is required 
when Einstein-Podolsky and Rosen
(EPR) discovered in 1935, certain nonlocal correlations in quantum
phenomenon.  They found that two systems, which are in an entangled
state, even if separated as far as you like from each other, retain
correlations, called EPR correlations, which do not decrease with
increasing separation.  Here by entangled state is meant those states
of the two systems which can not be written as product of their
individual systems in any basis whatsoever.  An example is the two
electrons in an spin singlet state.  Bohr felt this discovery to be
rather a bolt from the blue.  His response was basically that it does
not make sense to discuss parts of a combined system.

\newpage

\noindent {\large\bf 5. The Hidden variable program:}
\bigskip

We thus have seen that the assumption of the ``completeness of the
description by only the wavefunction of the system'' forces us to
renounce scientific realism, determinism, causlity, locality and loss
of unitary description of both the system and the apparatus.  It might
appear natural then to give up this assumption and entertain the
possibility of a more complete description of the state of the system
than that provided by the wavefunction alone by introducing additional
physical variables.  Such additional variable are now called the
``hidden variables'' though the terminology is not always a happy one.
To Bohr and Copenhagen school any such considerations were considered
an anathema and were ruled out of court.  They would have presumably
lessened the mystique of new discoveries.
\bigskip

Apart from the role played by the reigning Copenhagen
interpretation, it was a theorem, proved by the great mathematician
John von-Neumann, in 1932, which proved most discouraging to anybody
trying to follow a hidden variable program toward completion of
quantum mechanics.  According to this theorem any such completion
through hidden variables would not be able to reproduce all the
objective results of quantum mechanics.
\bigskip

David Bohm, nevertheless, published in 1954 a hidden variable theory,
which was not supposed to be possible by von-Neumann, for the
nonrelativistic quantum theory.  He proposed that the wavefunction together
with particle positions provide a valid completion of quantum theory.
The particle positions are the ``hidden variables'' in this Bohm's    
realistic causal interpretation of the quantum mechanics.  A similar proposal
had been made earlier by Louis de Broglie, who had earlier associated 
the concept of the waves for the particles, in 1927.  But under
criticism from Pauli and Einstein, he had given up this attempt.  Bohm
was able to deal with these early criticisms as well.  This
interpretation therefore is sometimes referred to as de-Broglie-Bohm
causal interpretation. We will mostly refer to it as Bohm's theory.
\bigskip

Under the spell of von-Neumann's theorem and of Copenhagen
itnerpretation, the work of Bohm also did not receive much attention.
The spell was only broken after John Bell started doing his important
work on foundations of quantum mechanics in 1966-67.  Bell analysed
the von-Neumann's proof, and since he had an explicit hidden variable
model of a spin one-half particle, he could pin-point an assumption in
von-Neumann's proof, which while looking mathematically nice, was not
necessarily physically required for the possible hidden variable
theories.  He also advocated Bohm's theory strongly.  Bell also
reformulated the theory.
\bigskip

\noindent {\large\bf 6. Constructing Bohm's theory:}
\bigskip

Bohm begins with the nonrelativistic Schr\"odinger equation for the
wavefunction for the case of a particle in a potential.  As we have
noted earlier the wavefunction is a complex function of space
coordinates and time.  Just as real numbers can be put in a one-to-one
correspondence with a line, the complex numbers can be put in a
one-to-one correspondence to an two dimensional Euclidean plane.  A
complex number can be specified by giving the distance of it's
corresponding point in the plane from the origin, ie ``modulus'', and
its ``phase'' ie the angle which the line from the origin to the
corresponding point make with one of the two perpendicular lines
(called $x$ and $y$ axis), say $x$ axis. Both the ``modulus'' and the
``phase'' of a complex number are real numbers.
\bigskip

We now rewrite the Schr\"odinger equation for time development.  The
complex wave function $\psi$ in terms of two equations, involving only
real quantities, for the time development of the square of its modulus
$R^2$ and the phase $\phi$.  Note that $\psi = R e^{i\phi}$, and $R^2 =
|\psi|^2$.  We also define action $S = \hbar \phi$ where $\hbar$ is
Planck's constant $h$ divided by $2\pi$.
\bigskip

The equation for the time evolution of the action looks quite similar
to Hamiltonian-Jacobi equation for the time development of action in 
classical dynamics except for an additional term, which we will call
``quantum potential'', $Q$.  It is given by $Q = -
\displaystyle{\hbar^2 \over 2m} \displaystyle{\nabla^2 R \over R}$ and 
formally vanishes as $\hbar \rightarrow 0$.  It adds to the potential
$V$ in which the particle was placed.  Effectively the potential which
the particle feels in Bohm's theory is $V + Q$.  Thus $S$ can be taken
as the ``quantum action''.
\bigskip

The equation for the time evolution of the square modulus $R^2 =
|\psi|^2$ has the form of a equation
of continuity for the density $R^2$ provided the momentum of the
particle is identified with the gradient of the action $S$, as is
natural in the Hamilton-Jacobi theory.  Madelung, in 1926, had tried
to identify the $R^2 \equiv |\psi|^2$ as the fluid density of electron
fluid in his hydrodynamical interpretation of quantum mechanics.  That
however was untenable as electrons were found to be localised objects
when they were detected and not spread out as in a fluid.  It was
however fully clear, after Born's work of 1926, that the value of
$|\psi|^2$ at a given location has to be interpreted as the
probability density of finding the quantum particles at that location.
Bohm also subscribed to it.
\bigskip

The identification of momentum with the gradient of the phase of the
wave function also leads to an expression for the velocity of the
particle since momentum is equal to mass time velocity.  We will refer
to it as the guidance equation for the particles.
\bigskip

The multiparticle generalisation of the above procedure is
straightforward.  We begin with the $N$-particle Schr\"odinger
equation and following the same procedure we find that particle
momenta are again given by the respective gradients of the action
function.  The Quantum potential now is given by the ratio of a sum of
$N$-particle Laplacians of $R$, each multiplied by a factor
$(-\hbar^2/2m)$ for the appropriate mass $m$, and divided by the $R$.
\bigskip

We now briefly recapitulate Bohm theory.  In Bohm's theory the basic
ontological entities are the wavefunction of the system and all the
particle positions. Both the wavefunction and particle positions obey
time evolution equations, ie Schr\"odinger equation for wavefunction
and the guidance equation for particle problems, are of first order in
time.  As a result once the wavefunction and particle positions are
given at an initial time they are determined at all later times.  The
trajectories are guided by the wavefunction.  It is therefore
sometimes referred to a ``pilot wave'' theory.  The wavefunction is
however not affected by the particle motion.
\bigskip

\noindent {\large\bf 7. Role of probability in Bohm theory:}
\bigskip

Bohm's theory is fully deterministic.  So where does the randomness in
quantum phenomenon come from?  It is taken, in Bohm theory, that we
are unable to control the particle positions precisely, so we are able
to prepare only that ensemble of particles in which the particle
positions of are distributed, at a given time, say $t=0$, a recording to
the probability distribution given by $|\psi (q,t = 0)|^2$.  We shall
refer this hypothesis as Bohm's ``quantum equilibrium hypothesis''.
Once this hyothesis is accepted Bohm's theory and standard quantum
mechanics lead to same observable consequence.  Once this initial
ensemble is prepared, then the laws of Bohmian dynamics make sure that
the particle positions of are distributed according the probability
distribution given $|\psi(q,t)|^2$ at later times.
\bigskip

The probability considerations thus appears in Bohm's theory in the
same way as they do in the classical statistical mechanics ie through
our ignorance of the precise initial conditions.  They are however not
intrinsic to the theory.
\bigskip

\noindent {\large\bf 8. Special Role of Particle Positions:}
\bigskip

It will be noticed that the particle positions play a rather special
role in the Bohm's theory.  It is conceptually independent of the
wavefunction and has it's own dynamical motion.  Since we unable to
produce particle ensembles, as we can not control the particle
position in it, other than those conforming to ``quantum equilibrium
hypothesis'', they are called hidden variables of the theory.  Further
in Bohm's theory it is assumed that they are, as Bohm and Hiley put
it, ``intrinsic and not
inherently dependent ..... on the overall context''.  They can be
measured without being changed.  In Bell's terminology they are
`beables' of the theory and not just `observables'.  In view of this,
all measurements are reducible, in Bohm's theory, to the pointer
readings of the measuring apparatus.
\bigskip

The particle momenta, given by the gradient of the action, depends on
the wavefunction of the system as a whole.  It is also a hidden
variable of the theory.  It is however not regarded as an intrinsic
property and is not a beable of the system.  A measurement does not reveal a
momentum value given by the Bohmian expression.      
\bigskip

\noindent {\large\bf 9. Waves or (/and) Particles:}
\bigskip

How does Bohm's theory view the particle or wave conundrum for quantum
objects like electron.  Bohm's theory associates both a position and
velocity, and therefore a trajectory, as well as a wavefunction to the
electrons.  So the simple answer is that electrons are not either a
particle or a wave but rather both a particle and a wave.
\bigskip

In the double slit experiment the electron trajectory goes through
only one of the slits but the electron wave, described by the
wavefunction, of course goes through both the slits.  This produces
the observed interference pattern on the screen.
\bigskip

If we wish to obtain the ``which way'' information ie to know as to
which particular slit any electron went through, we will have to put
electron track detectors near the two slits.  To discuss this new
situation we have to also include the detectors, along with electrons
and the slits, in our quantum description.  This discussion requires a
theory of system-detector interaction and will be dealt with later.
We will then see that getting this ``which way'' information destroys
the observed interference pattern.
\bigskip

\noindent {\large\bf 10. Nonlocality:}
\bigskip

The guidance equation for a particle velocity explicitly depends on
the gradient of phase of the wavefunction evaluated at the positions
of all the particles at that time and some of the other particles can
be quite far away.  This dependance of the particle velocities, on the
far away positions of other particle, is clearly nonlocal.  Same point
can be made through a consideration of the quantum potential which
also has a dependance on the position of other particles, some of
which could far away.  Bohm's theory is thus manifestly nonlocal.
What is the origin of this nonlocality in quantum mechanics.  As Bell
says, ``that the guiding wave, in the general case, propagates not in
ordinary three-space but in a multidimensional-configuration space is
the origin of the notorious `nonlocality' of quantum mechanics.  It is
a merit of the de Broglie-Bohm version to bring this out so explicitly
that it can not be ignored''.
\bigskip

The nonlocal dependence on the far away particle positions, of the
guidance equation for the motion of a particle, however disappears if
the wavefunction is separable in coordinates.  In general for all
two-particle wavefunction, which are entangled, nonlocal
Einstein-Podolsky-Rosen correlations will be there. They are easy to
understand in a natural way through Bohm's theory as it provides a
causal mechanism to generate them.
\bigskip

These nonlocal correlations however donot produce nonlocal
controllable effects.  So they can not be used for signalling
instantaneously and are thus from a physical point of view
comparatively benign.  This comes about since in quantum mechanics
such a signalling is not possible, and in view of Bohm's ``quantum
equilibrium'' hypothesis, all the observable consequences of Bohm's
theory agree with the standard quantum mechanics.
\bigskip

\noindent {\large\bf 11. Describing the ``measurements'':}
\bigskip

How does Bohm's theory cope with the notorious ``measurement'' problem
of the quantum mechanics?  For Bohm the measuring appartus is also to
be described by quantum mechanics.
\bigskip

Let initially the quantum system be in a definite state $i$, with
wavefunction $\psi(i)$, of the physical observable to be measured, and
let the measuring apparatus be in some fixed known base state with
it's pointer reading at $a_0$, with wavefunction $\phi(a_0)$ ie the
initial state of the round system is given by $\psi(i) \phi(a_0)$
interaction between the system and the apparatus causes the joint
system-apparatus state to evolve into the apparatus state to get
correlated with the system in the state $i$, so that at the completion
of the measurement, the Schr\"odinger unitary evolution of the joint
system leads to it's wavefunction becoming $\psi(i) \phi(a_i)$.  By
reading the pointer reading of the measuring apparatus to $a_i$, we
will conclude that the system was in the state $\psi(i)$.  If the
initial state of the physical system is a linear combination of
different states given by the normalised $\psi = c_1 \psi_1 + c_2
\psi_2 + \cdots$, that the initial joint state $\psi\phi(a_0)$ would
evolve to the joint state given by
\[
\psi = c_1 \psi_1 \phi(a_1) + c_2 \psi_2 \phi(a_2) + \cdots ,
\]
as the Schr\"odinger evolution is linear.
\bigskip

Now in view of the quantum equilibrium hypothesis, the configuration
of the system plus apparatus will be distributed according the
configuration probability density equal to $|\psi|^2$.  Now
\[
|\psi|^2 = |c_1 \psi_1 \phi(a_1)|^2 + |c_2\psi_2 \phi(a_2)|^2 + \cdots
 .
\]
We have here taken into account the fact that the different pointer
states, being macroscopically different, would have nonoverlapping
support in the configuration space of the apparatus ie $\phi(a_i)
\phi(a_j) = 0$ for $i \neq j$.  For the pointer reading to be equal to
$a_I$, the probability would be given by $|c_I|^2$.  This agrees with
Born's probability rule.  Further the system would be effectively in the state
$\psi(I)$.  We thus reproduce the results obtained from the ``collapse
of the wavefunction'' postulate of the standard quantum mechanics
without requiring any collapse of the wavefunction since the
wavefunction of the joint system $\psi$ does not collapse.
\bigskip

Let us call each $\psi(i)$ a channel for the system-particles.  After
the pointer reading is $a_I$, the system particles would be in the
channel $\psi(I)$.  Since their future particle motion in Bohm's
theory depends on their present positions, the only relevant part of
the wavefunction for it would be $\psi(I)$.  The other channels
$\psi(i)$ for $i \neq I$, are called empty waves.  They will continue
to evolve according to Schr\"odinger equation but are irrelevant for
the future motion of the system particles.
\bigskip

This discussion can be applied in a straightforward way to situation
of two slit experiments for the case when we position electron path
detectors near the two slits.  Bohm theory would reproduce the result
that in this case the interference pattern disappears.  In fact it was
not even necessary for us to point this out explicitly in view of our
demonstration above of the equivalence for all observable predictions
between Bohm's theory and standard quantum rules provided ``quantum
equilibrium'' hypothesis holds.
\bigskip

\noindent {\large\bf 12. Bohmian Mechanics:}
\bigskip

In Bohm's original formulation the modified Hamilton-Jacobi equation,
and the continity equation played an important role.  The momentum of
the particle was defined as in Hamilton-Jacobi theory.  Bohm regarded
particles moving under the influence of the forces just as in
Newtonian theory except that now they were subject an additional
Quantum force due the new quantum potential $Q$.  The Quantum
potential was used extensively to understand various quantum
phenomenon.  It served as a measure of deviation of the quantum
dynamics from the Newtonian one.  It was useful in many other
contexts.  In fact the textbooks of Holland, and of Bohm and Hiley on
Bohm's theory follow this approach.
\bigskip

Bohm was, of course, aware that his theory can be reformulated as a
first order theory by taking the Schr\"odinger equation for time
evolution of the wavefunction $\psi$ and the guidance equation for the
particle velocities specified in terms of wavefunction and it's
gradients.  This formulation was preferred by John Bell in his
presentation of theory.  It has been used D\"urr and his collaborators
extensively and they have named it Bohmian Mechanics.
\bigskip

Bohmian Mechanics appears to be a clearer and deeper formulation of the
theory.  Within this formulation one has been able to probe the nature
of quantum equilibrium hypothesis.  As we noted earlier if the
probability for the configuration $q$ is given $|\psi (q,t = 0)|^2$ at
some time, say $t=0$, then the distribution is given by
$|\psi(q,t)|^2$ for $t > 0$ ie the form of the distribution in terms
of the wavefunction does not change with time.  Thus assumed ``quantum
equilibrium'' has this attractive property which has been called
``equivariance'' by D\"urr et. al.  It has been shown later that it is 
the unique equivariant distribution which is a local functional of
$\psi$ by Goldstein et. al.  This concept of equivariance generalises
the concept of equilibrium distribution we come across in classical
statistical mechanics, e.g. Maxwellian distribution of particle
velocities in a gas.
\bigskip  

D\"urr, Goldstein and Zhangi tried to
argue that if deal with the wavefunction of the whole universe $\psi$,
then $|\psi|^2$ is a natural measure of probability for initial
configurations of the whole universe, which yields Born's rule for all
subsystems at a later time.  They argue that this measure is necessary
if there has to exist the notion of an effective wavefunction for the
subsystem. 
\bigskip

Now while we normally get the Maxwellian distribution in a gas, we can
concieve of situations, admittedly nonequilibrium ones, where it may
not be there e.g. by perturbing the thermal equilibrium of the gas.
The deviation from Maxwellian distribution, however, rapidly tend to
vanish.  Could it be that ``quantum equilibrium hypothesis'' in
Bohmian theory is of similar nature?  Valentini and Westman have
tried to argue that this is indeed quite plausible using an analogue
of classical coarse graining $H$-theorem of Boltzmann.  The
$H$-function defined by them is the
\[
H = \int dq \ \rho \ \ln(\rho/|\psi|^2)
\]
where $\rho$ is the arbitrary initial probability density in
configuration which tends to $|\psi|^2$ quickly.  This develops an
approach taken earlier by Bohm and Vigier.

\newpage

\noindent {\large\bf 13. Discontents with the Bohm theory:}
\bigskip

The most common objection against entertaining Bohm's realist
interpretation is since it has identical prediction to standard
quantum theory what is gained by introducing ``hidden variables''
referring to the positions of the particles in the theory.  If a
theory is nothing more than a set of calculational algorithisms for
predicting the result of the experiments then obviously nothing is
gained.  If the theory however is also supposed to provide an
understanding of the physical phenomenon then Bohm's theory definitely
does so better than the bare ``quantum rules''.  It gets rid of the
notorious ``measurement problem'' of standard quantum mechanics and
provides a unitary description of system and apparatus within the same
framework.
\bigskip

The position and momentum, are treated in a similar manner by the
standard quantum kinematics.  This feature is lost in the Bohmian
theory which gives particle position a special role in contrast to
momentum.  Of course the dynamics treats the position and momentum
asymmetrically e.g. the Hamiltonian is not symmetrical in the two.  So
it is not necessary for Bohm's theory to do so even if the standard
quantum kinematics does so.  Besides there is an attempt to write down
a version of Bohm's theory which actually treats the position and
momentum symmetrically.
\bigskip

Many people dread a return to the days of orderly classical physics
after having tasted the revolutionary fervour of the Copenhagen
interpretation. However Bohm's theory by no means does that.  The Bohm
theory, though sharply formulated as opposed to the fuzzy formulation
of Bohr and collaborators, does not return us to Newtonian mechanics.
The trajectories of the particles are very different in behavior.
Some times they are so far from Newtonian expectations that they have
been called surreal.  For example the electrons are at rest in the
bound states of a Hydrogen atom.  Bohm theory has also been
criticised, from the opposite side, for leading to such non-newtonian
trajectories.  The explicit nonlocality of Bohm theory also did not
endear it to people like Einstein.  This has however now, that we know
from experiments on Bell's inequality, is to be regarded rather a
virtue than as a defect.
\bigskip

Bell, however, found de Broglie-Bohm theory very instructive.  As he
advocated in 1982, ``should it not be taught, not as the only way, but
as an antidote to the prevailing complacency?  To show that vagueness,
subjectivity, and indeterminism, are not forced on us by experimental
facts, but by deliberate choice?''
\bigskip

\noindent {\large\bf 14. Future Prospects:}
\bigskip

\noindent 14.1 \underbar{Spin and Relativity}
\bigskip

Our discussion so far has been restricted spinless nonrelativistic
quantum mechanics.  Can it be extended to include spin and special
relativistic considerations.
\bigskip

Let us first discuss electrons which have a spin of one-half in the
units of $\hbar$.  We can use nonrelativistic Pauli equations, instead
of Schr\"odinger equation for spinless case, for this and it is
relatively easy to give a Bohm theory for this case.  This was done by
Bohm, Schiller and Tiomno.  It's interpretation is that of a particle
which is spinning rigidly.  One here defines a spin vector for each
electron. This faces some problems for the many electron problems.  A
more satisfactory approach is to regard nonrelativistic electrons as
described by 
the nonrelativistic limit of relativistic Dirac equation for an
electron.  Now the spin is not regarded as an inherent property in
addition to the particle velocity.  The spin effects arise due to an
extra term in the expression for the particle velocity itself.
\bigskip

Even though there are approaches to Dirac-Bohm theory using Dirac
equation for the relativic quantum mechanics, we do not discuss them
here.  We know that the formulation of relativistic quantum mechanics
without introducing quantum fields has not been a great success theory
irrespective of Bohm theory.
\bigskip

\noindent 14.2 \underbar{Field Theory}
\bigskip

We shall now discuss whether Bohm like formulation can be given for
quantum field theories.  It was believed for a long time that it can
not be done.  This was so despite the fact that Bohm himself had
applied his theory to electromagnetic fields in an appendix to his
1952 papers.  Bohm and Hiley applied it scalar fields later.  Thus
application to Bosonic fields seems to present no undue
difficulties. Here one introduces ``field beables''.  Bosons,
e.g. photons, do not have a trajectory.
\bigskip

It is however true that an application of the Bohm theory for the
fermionic fields waited till Bell's attempt to introduce fermionic
particle beables for them in 1984.  The attempts using fermionic field
beables do not seem to be very successful.  Bell's work was done for a
``lattice'' cutoff model of the field theory and was stochastic in
nature.  He however suspected that the ``stochastic element introduced
here goes away in some sense in the continuum limit'' ie when lattice
cut off's are removed.  In a continuum model later developed by Colin
et. al., this seems to be the case.
\bigskip

D\"urr et. al. have developed a Bohmian mechanics version of continuum
field theory which they have called Bell-type quantum field theories.
They associate particle ontology with both bosonic and fermionic
fields.  The interaction part of the Hamiltonian is associated with
jump-like stochastic processes like the ``particle-antiparticle pair''
creation or annihilation.  
\bigskip

A common feature of the work on Bohmian field theory is lack of
manifest Lorentz covariance.  From what one has said clearly much work
remains to be done in this area.
\bigskip

\noindent 14.3 \underbar{New Problems}
\bigskip

We had emphasised earlier that all the observable consequences of the
standard quantum mechanics are reproduced by the Bohm theory provided
quantum equilibrium hypothesis is accepted.  But there is a prospect
that since Bohm theory is a sharper formulation of the quantum
mechanics, it allows us to formulate new problems which can not even
be formulated in the old language of the standard quantum mechanics.
An example of such a problem is ``How much time does a particle spends
in the potential barrier?''  Bohm theory, having trajectories, has a
definite answer while the standard quantum mechanics does not even
allow us to ask the question.  If the feasible experiments can be
devised for measuring these ``dwell-times'', then clearly one has some
thing to look for.
\bigskip

A remoter possibility is follows. Suppose a future technology allows
us to prepare ensembles of particle, which are not having ``quantum
ensemble'' distribution, then we should observe derivations from the
predictions of ordinary quantum mechanics.  Of course, not all
possible deviations from ``quantum equilibrium'' hypothesis can occur
as some of them would lead to instantaneous signalling which would
violate special theory of relativity.  A possibility has also been
considered by Valentini that the quantum equilibrium was not
established at the time of big bang of the universe, and if so, it
would have observable consequences in that there would be corrections
to the usual inflationary model predictions for cosmic microwave
background and super-Hubble field correlations and relic
nonequilibrium particles.
\bigskip

\noindent {\large\bf 15. Mathematical Notes}
\bigskip

Here we collect some mathematical material relevant to Bohm's theory.
This section can be skipped by nonphysicists.  For physicists however
this section would add to their deeper understanding and enjoyment.
\bigskip

\noindent 15.1 \underbar{Equations for de Broglie-Bohm's causal
theory}
\bigskip

We first consider a single particle, with mass $m$, moving in a
potential $V(\vec r)$.  The time $t$, development of the wavefunction
$\psi(\vec r)$ of the system, with Hamiltonian $H$, where
\be
H = {(\vec p)^2 \over 2m} + V(\vec r),
\label{one}
\ee
is given by
\be
i\hbar {\partial\psi \over \partial t} = - {\hbar^2 \over 2m} \nabla^2
\psi + V(\vec r)\psi.
\label{two}
\ee
Let us rewrite the wavefunction in it's polar decomposition given by
\be
\psi = R e^{iS/\hbar}
\label{three}
\ee
where $R$ and $S$ are real functions of $\vec r$.  Substituting the
decomposition (\ref{three}) in eqn.(\ref{two}) and separating out the
real and imaginary parts of the equation we obtain
\be
{\partial R \over \partial t} + {1 \over 2m} \left[R \nabla^2 S + 2
\nabla R \cdot \nabla S\right] = 0 
\label{four}
\ee
and
\be
{\partial S \over \partial t} + {(\nabla S)^2 \over 2m} + V(x) -
{\hbar^2 \over 2m} {\nabla^2 R \over R} = 0.
\label{five}
\ee

We now define
\be
P(\vec r,t) = |\psi(\vec r,t)|^2 = |R(\vec r,t)|^2,
\label{six}
\ee
\be
m \vec v(\vec r,t) = \nabla S(\vec r,t),
\label{seven}
\ee
and
\be
Q(\vec r,t) = - {\hbar^2 \over 2m} {\nabla^2 R(\vec r,t) \over R(\vec
r,t)}. 
\label{eight}
\ee
Using these definitions we can rewrite eqns.(\ref{four}) and
(\ref{five}) in more suggestive forms
\be
{\partial P \over \partial t} + \vec \nabla(P \vec v) = 0
\label{nine}
\ee
\be
{\partial S \over \partial t} + {(\nabla S)^2 \over 2m} + V + Q = 0.
\label{ten}
\ee
The first of these equations is the continuity of equation for density
$P$ with an associated current density $P \vec v$. The second of these
equations is of the form of Hamiltonian-Jacobi equation in Newtonian
dynamics for a mass $m$ particle moving in the potential $V + Q$.  The
$Q$ is thus an added potential of quantum origin and is referred to as
Quantum potential.
\bigskip

The de Broglie-Bohm theory also introduces the particle positon $\vec
q (t)$ as the extra variable needed to describe the system fully in
addition to the wavefunction $\psi(\vec r,t)$.  We further make the
identification of the particle velocity, $d\vec q(t)/dt$ with $\vec
v(\vec r,t)$, defined above in equation (\ref{seven}) at $(\vec r,t)
\equiv (\vec q(t),t)$ ie
\be
{d\vec q \over dt} = \vec v(\vec q,t) = {1 \over m} \nabla S(\vec
r,t)\bigg|_{(\vec r,t)=(\vec q,t)},
\label{eleven}
\ee
since it is natural in Hamilton-Jacobi theory to have particle
momentum $\vec p(t)$
\be
\vec p(t) = m {d \vec q \over dt} = \nabla S(\vec r,t)\bigg|_{(\vec
r,t)=(\vec q,t)}.
\label{twelve}
\ee
With these definitions, it can be shown that
\be
\left[{\partial \over \partial t} + {1 \over m} \nabla S \cdot
\nabla\right] \nabla S = -\nabla(V + Q).
\label{thirteen}
\ee
This can be rewritten as
\be
{d \over dt} \vec p(t) = -\nabla(V + Q)
\label{fourteen}
\ee
where
\be
{d \over dt} = {\partial \over \partial t} + \vec v \cdot \vec\nabla, 
\label{fifteen}
\ee
which is the Newtonian equation of motion for a particle of mass $m$
in the potential $V + Q$.  This justifies the name ``Quantum
Potential'' for $Q$ for it plays the same role as potential $V$.
\bigskip

Following Born, Bohm also identifies the probability density
$\rho(\vec q,t)$ for the particle positions $\vec q(t)$, as follows,
\be
\rho(\vec q,t) = P(\vec q,t) = |\psi(\vec q,t)|^2.
\label{sixteen}
\ee
The ``quantum equilibrium'' hypothesis is that
\be
\rho(q,t=0) = |\psi(\vec q,t=0)|^2.
\label{seventeen}
\ee
It then follows from de Broglie-Bohm equations of motion that
\be
\rho(q,t) = |\psi(\vec q,t)|^2.
\label{eighteen}
\ee

The $N$-particle generalisation is straightforward.  One has now to
have $N$ trajectory functions $\vec q_1(t),\vec q_2(t),\cdots,\vec
q_N(t)$ in addition to the wavefunction $\psi(\vec r_1,\vec
r_2,\cdots,\vec r_N;t)$ for a complete description of the system.  We
now have, apart from the usual Schr\"odinger equation for the $N$
particle system
\be
m_i {d \vec q_i \over dt} = \nabla_i S\bigg|_{\vec r_i = q_i},
\label{ninteen}
\ee
where $m_i$ is the mass of the $i$-th particle.  The Quantum potential
$Q$ is given by 
\be
Q = - {\hbar^2 \over 2} \left({1 \over m} {\nabla^2_1 R \over R} +
{1 \over m_2} {\nabla^2_2 R \over R} + \cdots + {1 \over m_N}
{\nabla^2_N R \over R}\right)\bigg|_{\vec r_i - q_i}.
\label{twenty}
\ee
\bigskip

\noindent 15.2 \underbar{Equations for Bohmian Mechanics}
\bigskip

Newtonian equations of motion, eqn. (\ref{fourteen}), are second order
in time as they specify the particle accelerations $d\vec p/dt$ or
$d^2\vec q/dt^2$.  The equations of Bohmian mechanics are first order
in time.  They are
\begin{enumerate}
\item[{(i)}] Schr\"odinger equation for the wavefunction $\psi(\vec
r_1,\vec r_2,\cdots,\vec r_N,t)$ for a $N$-particle system, with
$i$-th particle having mass $m_i$, and moving in a potential $V(\vec
r_1,\vec r_2,\cdots,\vec r_N)$ given by
\be
i\hbar {\partial \psi \over \partial t} = -\left({\hbar^2 \over 2m_1}
\nabla^2_1 + {\hbar^2 \over 2m_2} \nabla^2_2 + \cdots + {\hbar^2 \over
2m_N} \nabla^2_N\right)\psi + V(\vec r_1,\vec r_2,\cdots,\vec
r_N)\psi, 
\label{twentyone}
\ee
and
\item[{(ii)}] particle guidance equations
\be
m_i {d\vec q_i(t) \over dt} = {\hbar \over 2i} \left({\nabla \psi
\over \psi} - {\nabla \psi^\star \over
\psi^\star}\right)\bigg|_{\vec r_i = \vec q_i}.
\label{twentytwo}
\ee
The quantum equilibrium hypothesis is same as before.
\end{enumerate}
\bigskip

\noindent {\large\bf 16. Bibiliographical Notes}
\bigskip

\begin{enumerate}
\item[{1.}] For a somewhat more elaborate account of the classical
physics and it's world view, see \\ Virendra Singh, Scientific Realism
and Classical Physics, TIFR preprint, TIFR/TH/08-17, (2008).
\item[{2.}] For a historial account of Quantum mechanics, see \\
M. Jammer, \underbar{The Conceptual Development of Quantum Mechanics,
New York, 1966} and \underbar{The Philosophy of Quantum Mechanics},
New York, 1974, \\ J. Mehra and H. Rechenberg, \underbar{The
Historical Development of Quantum Mechanics}, Springer, 1982-$\cdots$
. \\ M. Beller, \underbar{Quantum Dialogue}, Chicago, 1999. 
\item[{3.}] For the basic formalism of quantum mechanics, see e.g., \\
P.A.M. Dirac, \underbar{The Principles of Quantum mechanics}, (fourth
edition), Oxford, 1958. \\ J. von-Neumann, \underbar{Mathematische
Grundlagen der Quanten Mechanik}, Berlin, 1932, \\
\underbar{Mathematical Foundations of Quantum Mechanics} (english
trans. by R.T. Beyer), Princeton, 1955 \\ A. Bohm, \underbar{Quantum
Mechanics: Foundations and Applications}, (3rd ed.), New York,
2001. 
\item[{4.}] (i) A number of important basic sources on quantum
mechanics are available in, \\ J.A. Wheeler and W.H. Zurek,
\underbar{Quantum Theory and Measurement}, Princeton, 1983. \\ (ii)
The far reaching writing of John Bell are collected in \\ J.S. Bell,
\underbar{Speakable and Unspeakable in Quantum mechanics}, Cambridge,
1987. \\ These contain his papers on hidden variables in quantum
mechanics, E.P.R. correlations, de Broglie-Bohm theory and other
fundamental aspects on quantum mechanics.
\item[{5.}] The Copenhagen interpretation is discussed in many books.
For Bohr's views we refer to \\ N. Bohr, \underbar{Atomic Theory and
the Description of Nature}, Cambridge, 1934; \\ N. Bohr, \underbar{Atomic
Physics and Human Knowledge}, New York, 1958; \\ N. Bohr,
\underbar{Essays 1958/1962 on Atomic Physics and Human Knowledge}, New
York, 1963. \\ For the views of Heisenberg, we refer to \\
W. Heisenberg, \underbar{The Physical Principles of the Quantum
Theory}, New York, (1930); \\ W. Heisenberg, \underbar{Physics and
Philosophy}, New York, 1958. \\ von-Neumann's theory of measurement in
quantum mechanics is contained in his book cited earlier. \\ There is
also the monograph, \\ D. Murdoch, \underbar{Niels Bohr's Philosophy
of Physics}, Cambridge, 1987. \\ We also recommend the following
textbooks, \\ L.D. Landau and E.M. Lifshitz, \underbar{Quantum
Mechanics}, New York, 1958; \\ K. Gottfried, \underbar{Quantum
mechanics}, New York, 1966; \\ D. Bohm, \underbar{Quantum Theory}, New
York, 1951. \\ Surprisingly the textbook on quantum mechanics, written
by Bohm somewhat before he developed his causal interpretation, also
takes Copenhagen point of view.  See also \\ H.P. Stapp,
Am. Jour. Phys. \underbar{40}, 1098 (1972).
\item[{6.}] The best description of the double slit experiment is
given in,  \\ R.P. Feynman, \underbar{The Feynman Lectures on Physics,
Vol. 3} (Quantum Mechanics), New York (1966); Narosa indian eighth
reprint, 1992. \\ See also, \\ J.A. Wheeler, The `past' and `delayed
choice' double slit experiment, in \\ H.R. Harlow (ed.),
\underbar{Mathematical Foundations of Quantum Mechanics}, New York,
1960. 
\item[{7.}] The original E.P.R. papers appeared in \\ A. Einstein,
B. Podolsky and N. Rosen, Phys. Rev., \underbar{47}, 777 (1935); \\
Bohr's response appeared in \\ N. Bohr, Phys. Rev. \underbar{48}, 696
(1935) \\ D. Bohm's reformulation, using two spin one-half particles,
was given in his quantum theory textbook of 1951. \\ Bell's original
paper on Bell's inequality appeared in \\ J. Bell, Physics
\underbar{1}, 195 (1964). \\ All of these papers have been reproduced
in the ``Wheeler-Zureck'' collection referred to earlier.  Bell's paper
is of course available in Bell's book cited earlier. 
\item[{8.}] On the ``hidden variable'' we refer to the following: \\
(i) John von-Neumanns impossibility proof against the existence of
hidden variables in quantum mechanics is contained in his book cited
earlier. \\ (ii) J. Bell's original paper on hidden variables appeared
in \\ J. Bell, Rev. Mod. Phys., \underbar{38}, 447 (1966). \\ (iii)
For a recent review, we refer to \\ Virendra Singh, Hidden Variables,
Non-contextuality and Einstein Locality, arXive: quant-ph/0507182 V2
(2005).  To appear in an ICPR volume edited by P.K. Sengupta.
\item[{9.}] (i) The original papers on de Broglie-Bohm's theory, which
came to be known as ``causal interpretation of quantum mechanics''
are \\ L. de Broglie, J. Physique, 6th series, \underbar{8}m 225
(1927); \\ D. Bohm, Phys. Rev. \underbar{85}, 166 (1952); \\ D. Bohm,
Phys. Rev. \underbar{85}, 180 (1952); \\ (ii) For monographs on Bohm's
theory, see \\ D. Bohm and B.J. Hiley, \underbar{The Undivided
Universe: An ontological interpretation of} \\ \underbar{quantum
theory}, London, 
1993; \\ P.R. Holland; \underbar{The Quantum Theory of Motion: An
account of the de Broglie-Bohm} \\ \underbar{causal interpretation of quantum
mechanics}, Cambridge, 1993. \\ Both the monographs emphasise the
``quantum potential'' approach following Bohm. \\ (iii) For the
Bohmian Mechanics version, which was also favoured by Bell, see \\
K. Berndl, M. Daumer, D. D\"urr, S. Goldstein and N. Zhangi, A survey
of Bohmian Mechanics, Nuovo Cimento \underbar{110B}, 737 (1995)
$\equiv$ arXive. quant-ph/9504010; \\ D. D\"urr, S. Goldstein and
N. Zhangi, J. Stat. Phys. 67, 843 (1992) $\equiv$ quant-ph/0308039 \\
S. Goldstein, \underbar{Bohmian Mechanics}, in \underbar{Stanford 
Encyclopdia of Philosophy} ed. by E.N. Zalta, available on web at
http://plato.stanford.edu/archives/win2002/entries/qm-bohm/ . \\
R. Tumulka, Am. J. Phys. \underbar{72}, 1220 (2004). \\ At a more
popular level, see \\ D.J. Albert, Sc.Am. p.58-67 (May 1994). \\ (iv)
For two, rather illuminating accounts emphasising alternative possible
scenarios of the development of quantum theory in which de Broglie
Bohm theory rather than the Copenhagen interpretation had emerged as
the dominant view point, see \\ J.T. Cushing, \underbar{Quantum
Mechanics, Historical contingency and the Copenhagen Hegemony},
Chicago, 1994; \\ H. Nikoli\'c, Am. J. Phys. \underbar{76}, 143
(2008). \\ (v) For appraisals, see \\ J.T. Cushing, A. Fine and
S. Goldstein (ed.): \underbar{Bohmian Mechanics and Quantum} \\
\underbar{Theory: an appraisal}, Dordrecht (1996); \\ O. Passon, why
isn't every physicist: Bohmian?, arXive, quant-ph/0412119v2 (2005).
\item[{10.}] For discussions of the origins of the ``quantum
equilibrium'' hypothesis within Bohmian theory see \\ D. Bohm,
Phys. Rev. \underbar{89}, 458 (1953); \\ D. Bohm and J.P. Vigier,
Phys. Rev. \underbar{96}, 208 (1954); \\ D. D\"urr, S. Goldstein and
N. Zhangi, J. Stat. Phys. 67, 843 (1992) $\equiv$ arXive:
quant-ph/0308039; \\ D. D\"urr, S. Goldstein and N. Zhangi,
Found. Phys. \underbar{23}, 721 (1993); \\ A. Valentini and
H. Westman, Proc. Roy. Soc. A\underbar{461}, 253 (2005) \\
S. Goldstein and W. Struyve, On the Uniqueness of quantum equilibrium
in Bohmian mechanics, arXive: 0703.3070v1 (2007).
\item[{11.}] (i) For an attempt to treat position and momentum
formally symmetically see, \\ S.M. Roy and V. Singh, Modern Physics
Letters \underbar{A10}, 709 (1995); \\ V. Singh, Causal quantum
mechanics, Pramana \underbar{49}, 5 (1997); \\ S.M. Roy and V. Singh,
Phys. Letters \underbar{A255}, 201 (1999). \\ (ii) On surreal
trajectories objection, see \\ B.G. Englert, M.O. Scully,
G. S\"ussmann and H. Walther, Z. Naturforsch. \underbar{47a}, 1175
(1992); \\ M.O. Scully, Phys. Scr. \underbar{76}, 41 (1998). \\ For
some reactions see, \\ C. Dewdney, L. Hardy and E.J. Squires,
Phys. Lett., \underbar{A184}, 6 (1993) \\ D. D\"urr, W. Fusseder,
S. Goldstein and N. Zanghi, Z. Naturforsch. \underbar{48a}, 1161
(1993); see also Z. Naturforsch. \underbar{48a}, 1163 (1993); \\
B.J. Hiley, R.E. Collaghan and O.J.E. Maroney, arXive:
quant-ph/0010020 (2000).
\item[{12}] (i) For a treatment of spin 1/2 nonrelativistic particles,
see \\ D. Bohm, R. Schiller and J. Tiomno, Nuovo
Cim. Supp. \underbar{1}, 48 (1955); \\ EPR spin correlatins have been
discussed using this model by \\ C. Dewdney, P.. Holland and
A. Kyprianidis, J. Phys. \underbar{A20}, 4717 (1987); \\ For a further
discussion of the spin problem in Bohm see the books by Bohm and Hiley
and by Holland referred to earlier. \\ (ii) On Bohmian field theory
see \\ D. Bohm, Phys. Rev. \underbar{85}, 180 (1952) Appendix A; \\
D. Bohm and B. Hiley, Found. Phys. \underbar{14}, 270 (1984); \\
J. Bell, Beables for quantum field theory, CERN-Th-4035/ 84, (1984)
$\equiv$ Phys. Reports \underbar{137}, 49 (1986); \\ P.R. Holland,
Phys. Reports, \underbar{224}, 95 (1993), \\ S. Colin,
Phys. Lett. A\underbar{317}, 349 (2003); \\
D. D\"urr, S. Goldstein,
R. Tumulka and N. Zanghi, J. Phys. A\underbar{38}, R1 (2005), and
references therein. \\ S. Colin and W. Struyve,
J. Phys. A\underbar{40}, 7309 (2007); \\ W. Struyve and H. Westman, A
minimalist pilot wave model for quantum electrodynamics, arXive:
0707.3487v2 (2007): \\
(iii) For some problem involving tunnelling
time etc., see \\ C.R. Leavens, Solid State Comm. \underbar{76}, 253
(1990); \\ A.S. Majumdar and D. Home, Phys. Lett. \underbar{A296}, 76
(2002). \\ S. Kreidel, S. Gr\"ube and H.G. Embacher,
J. Phys. \underbar{A36}, 8851 (2003); \\ (iv) The suggestion about a
possible non quantum equilibrium scenario is contained in, \\
A. Valentini, De Broglie-Bohm prediction of Quantum violations for
cosmological super Hubble modes, arXive: 0804.4656 (2008). \\ (v) For
a discussion of relationships and comparisons of the Bohm's theory and
other interpretations of quantum theory, we refer to \\ S. Goldstein,
Physics Today, p. 42 (March, 1998) and p. 38 (April 1998); \\
R.B. Griffiths, Bohmian Mechanics and consistent histories, arXive:
quant-ph/9902059v2 (1999); \\ J.B. Hartle, Bohmian Histories and
Decoherent Histories, arXive: quant-ph/0209104v4 (2004); \\ Virendra
Singh, Quantum Mechanics and Reality; arXive: quant-ph/0412148 (2004),
to appear in a ICPR volume edited by B.V. Sreekantan; \\ V. Allori,
S. Goldstein, R. Tumulka and N. Zanghi, On the common structure of
Bohmian Mechanics and the Ghirardi-Rimini-Weber Theory, arXive:
quant-ph/0603027v4 (2007); \\ P.J. Lewis, Brit. J. Phil. Sci.,
\underbar{58}, 787 (2007). 
\end{enumerate}

\end{document}